\documentclass[a4paper,10pt,twocolumn,aps,showpacs,pra]{revtex4-1}
\usepackage{graphicx}% Include figure files
\usepackage{dcolumn}% Align table columns on decimal point
\usepackage{bm}% bold math
\usepackage{slashed}
\usepackage{amssymb}
\usepackage{amsmath}
\usepackage{amsthm}
\usepackage{color}

\newcommand{\ket}[1]{| #1 \rangle}

\newcommand{\ClebschG}[6]{\left[{#1\atop#2}{#3\atop#4}{#5\atop#6}\right]}
\newcommand{\CG}[6]{\left[{#1\atop#2}{#3\atop#4}{#5\atop#6}\right]}

\newcommand{\SixJ}[6]{\left\{{#1\atop#2}{#3\atop#4}{#5\atop#6}\right\}}

\begin{document}

\title{Negativity in the Generalized Valence Bond Solid State}

\author{Raul A. Santos$^{1}$ and V. Korepin$^{2}$}

\affiliation{$^{1}$Department of Condensed Matter Physics, Weizmann Institute of Science, Rehovot 76100, Israel}
\affiliation{$^{2}$C.N. Yang Institute for Theoretical Physics, Stony Brook University, Stony Brook, NY 11794-3840, US}
  
\begin{abstract}
Using a graphical presentation of the spin $S$ one dimensional Valence Bond Solid (VBS) state, based on the 
representation theory of the $SU(2)$ Lie-algebra of spins, we compute the spectrum of a mixed state reduced density 
matrix. This mixed state of two blocks of spins $A$ and $B$
is obtained by tracing out the spins outside $A$ and $B$, in the pure VBS state density matrix. We find in particular that the 
negativity of the mixed state is non-zero only for adjacent subsystems. The method introduced here can be generalized
to the computation of entanglement properties in Levin-Wen models, that possess a similar algebraic structure to the 
VBS state in the groundstate.
\end{abstract}

%\pacs{03.67.-a, 05.30.-d, 03.67.Mn}

\maketitle

\section{Introduction}

The use of quantum information concepts in the study and characterization of many-body systems
has proved highly fruitful \cite{Laflorencie2015}. A clear example of such concepts is the entanglement entropy, which became 
an standard tool to characterize gapped, topological and critical phases \cite{Korepin2004,Eisert2010,Kitaev2006,Levin2006}.
Despite its usefulness, the entanglement entropy does not fully capture the entanglement in multipartite systems.
In systems with more than two components, the state of any two subsystems is described in general by a mixed state.
In these systems, a measure of entanglement that generalizes the Perez-Horodecki criterion \cite{Peres1996} 
has been proposed. This measure, called negativity \cite{Vidal2002} is based on the 
 partial positive transpose (PPT). For a separable state, the partial transpose
density matrix (PTDM) is still a density matrix i.e. all its non-vanishing eigenvalues are positive and add to one. In contrast,
if a state is not separable, then its PTDM has negative eigenvalues \cite{Peres1996}. The
sum of the negative eigenvalues of the PTDM is the negativity ${\rm Neg}(\rho)$. This quantity is a good measure of
entanglement in the sense that it satisfies the following four fundamental criteria \cite{Vidal2002};
\begin{itemize}
\item i) ${\rm Neg}(\rho)\geq 0$, being zero just for unentangled states.
\item ii) ${\rm Neg}(U_{AB}\rho_{AB}U^\dagger_{AB})={\rm Neg}(\rho)$ for any unitary $U_{AB}=U_A\otimes U_B$.
\item iii) ${\rm Neg}(\rho)$ does not increase under Local Operations and Classical Communication (LOCC) or post-selection.
\item iv) ${\rm Neg}(\rho)$ is convex, i.e. for $p_i\leq 0$ and $\sum_ip_i$, $\sum_i{p_i}{\rm Neg}(\rho_i)\geq{\rm Neg}(\sum_i{p_i}\rho_i)$, meaning that it decreases
under discarding information.
\end{itemize}

In systems with a small Hilbert space, the computation of the negativity can be done directly. In many-body systems the sole
computation of this entanglement measure becomes challenging. Noteworthy progress has been achieved in systems with bosons
\cite{Eisler2014}, free fermions \cite{Eisler2015,Chang2016}, conformally invariant systems in and out of equilibrium 
\cite{Calabrese2013,Calabrese2015,Wen2015}, topological \cite{Lee2013,Castelnovo2013} and spin systems 
\cite{Santos2011,Santos2012,Fagotti2012}. 

In this article we present exact results for the negativity of a series of one dimensional interacting spin $S$ systems, whose
groundstate generalize the valence bond solid (VBS) state. The prime example of this type of systems is given by the 
Affleck, Lieb, Kennedy and Tasaki (AKLT) model. This model represents an example of a gapped spin one system, representative of
the Haldane phase \cite{Haldane1983,AKLT_Short,AKLT_LONG}. In the condensed matter community this model is interesting because it 
realizes a Symmetry Protected Topological (STP) phase \cite{Chen2011,Pollmann2012}. For open boundary conditions, the system exhibits fractionalized 
edge modes localized on the edge of the system, and a ground state degeneracy. In the quantum information community, this model 
has also received attention as a possible resource for measurement based quantum computation \cite{Miyake2011,Wei2011,Wei2014}.

The generalized model consists on integer spin $S$ particles arranged on a line. The (frustration free) Hamiltonian is a sum of 
projectors into total spin $S^{\rm tot}_{i,i+1}=S_{i} +S_{i+1}>S$ of consecutive particles. For $S=1$ it reduces to the usual AKLT model \cite{Kirillov1990}.
By tracing subsections of the chain, the reduced density matrix $\rho_{AB}$ of the mixed system of two intervals is obtained.
The method presented here to obtain the negativity relies heavily on the representation theory of the $SU(2)$ algebra.
We first write the ground state in a matrix product state (MPS) form, in terms of Clebsch-Gordan coefficients. This allows to 
represent the groundstate partial density matrix as a state sum in a quasi one dimensional lattice, where each link of the
lattice has associated a $SU(2)$ representation label. Using $F$-moves 
(matrices of change of basis between different but equivalent fusion orderings) and the symmetries of the Clebsch-Gordan 
coefficients, we can diagonalize $\rho_{AB}$ and its partial transpose $\rho^{T_A}_{AB}$.

This paper is organized as follows. In section II we introduce the generalized spin $S$ VBS model, it's MPS
representation and its graphical presentation based on the representation theory of the $SU(2)$ Lie algebra. In section III we 
define the reduced density matrix to study by tracing blocks of spins. We then compute the eigenvalues of such operator
by transforming the density matrix into a standard basis in the tensor representation. In section IV we analyze the 
eigenvalues of the partial transposed density matrix and compute the negativity. Lastly, in section V we present our conclusions.

\section{Spin S VBS state}

Let's consider the generalization of the AKLT model for general integer spin $S$ particles, with Hamiltonian
\begin{equation}\label{hamiltonian}
	\mathcal{H} = \sum_{i=1}^L h_{i,i+1} = \sum_{i=1}^L \sum_{s=S+1}^{2S} \Pi_s(i,i+1),
\end{equation}
where $\Pi_s(i,i+1)$ is a projector onto the subspace spanned by the $s$-multiplet formed by spins at $i$ and $i+1$.
This Hamiltonian is positive definite.
As in the AKLT case, the groundstate can be found exactly by considering each spin $S$ particle as the result of a projection
onto the symmetric subspace of two spin $S/2$ particles. Then, each virtual particle of spin $S/2$ is antisymmetrized with its 
nearest neighbor into the singlet state. The presence of this singlet between consecutive particles prevents the
formation of total spin $S_{i,i+1}=S_i+S_{i+1}$ larger than $S$. Repeating this procedure with every particle in the chain, we
obtain a state that is annihilated by the Hamiltonian (\ref{hamiltonian}). It corresponds to the groundstate of 
(\ref{hamiltonian}) with eigenvalue zero.

Let us now write down the MPS representation of the spin $S$ VBS state. 
General boundary conditions can be implemented by boundary tensors $u_1,u_L$
\begin{equation}\label{mps0}
	|\mathcal{G}\rangle = u_1^\dagger\mathbf{g}_1\mathbf{g}_2\dots\mathbf{g}_Lu_L,
\end{equation}
For a periodic chain of $L$ spins we have 
\begin{equation}\label{mps}
	|\mathcal{G}\rangle = {\rm tr}(\mathbf{g}_1\mathbf{g}_2\dots\mathbf{g}_L),
\end{equation}
where ${\bf g}_i$ are $(S+1)\times(S+1)$ matrices. The trace here is done over the auxiliary matrix space. The elements of 
$\mathbf{g}_i$ and its dual $\bar{\mathbf{g}}_i$ are state vectors:
\begin{eqnarray}\label{def_g}
 (\mathbf{g}_i)_{ab}&=&\sum_{m}{\ClebschG{\frac{S}{2}}{a}{\frac{S}{2}}{-b}{S}{m}}(-1)^b\ \ket{S,m}_i, \\
	(\bar{\mathbf{g}}_i)_{ab}&=&\sum_{m}\ClebschG{\frac{S}{2}}{a}{\frac{S}{2}}{-b}{S}{m}(-1)^b\langle{S,m}|_i,
\end{eqnarray}
with $|S,m\rangle$ a state of total spin $S$ and $S_z$ spin projection $m$.
Here $\ClebschG{S_1}{m_1}{S_2}{m_2}{J}{m}$ is the Clebsch-Gordan coefficient of the change of basis
between the tensor product of states with definite spin and $S_z$ projection $\ket{S_1,m_1}$ and $\ket{S_2,m_2}$ and the state
with definite {\it total} spin and $S^{\rm total}_z$ projection, i.e.
\begin{equation}
 |J,m\rangle=\sum_{m_1,m_2}\ClebschG{S_1}{m_1}{S_2}{m_2}{J}{m}\ket{S_1,m_1}\otimes \ket{S_2,m_2}.
\end{equation}
The state $\ket{\mathcal{G}}$ is annihilated by the generalized AKLT Hamiltonian (\ref{hamiltonian})
with periodic boundary conditions.

Given the symmetry under translations, the groundstate correlation functions are completely determined by the
so called transfer matrix \cite{Santos2012b}. The transfer matrix for this state is given by 
$G_{ab}^{cd}= (\bar{\mathbf{g}}_i)_{cd}(\mathbf{g}_i)_{ab}$, taking the inner products of the spin states. Explicitly
\begin{equation}\label{TM_1}
 G_{ab}^{cd}=\sum_m\CG{\frac{S}{2}}{a}{\frac{S}{2}}{-b}{S}{m}\CG{\frac{S}{2}}{c}{\frac{S}{2}}{-d}{S}{m}(-1)^{S+b+d}.
\end{equation}
Using the recoupling of Clebsch-Gordan coefficients \cite{Santos2012b}, (see Appendix \ref{appendixB}), the transfer matrix can be written in the form
\begin{equation}\label{TM_2}
 G_{ab}^{cd}=\sum_{j,m}\lambda_j\CG{\frac{S}{2}}{-a}{\frac{S}{2}}{c}{j}{m}\CG{\frac{S}{2}}{-b}{\frac{S}{2}}{d}{j}{m}(-1)^{a+b},
\end{equation}
where
\begin{equation}\label{lambdaj}
\lambda_j=(-1)^{j}\frac{(S!)^2(S+1)}{(S-j)!(S+j+1)!}.
\end{equation}

\subsection*{Diagrammatic presentation}

\begin{figure}[h!]
		 \includegraphics[width=\linewidth]{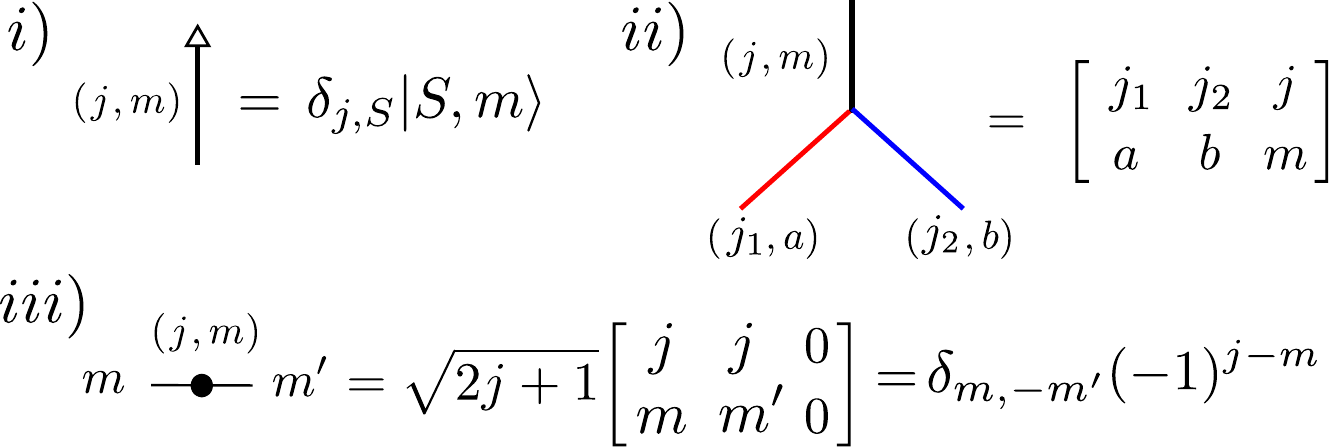}
	\caption{(color online) $i)$ A north directed arrow represents a ket state vector. An arrow directed south represents a bra (dual state
	vector). $ii)$ A Clebsch-Gordan coefficient corresponds to a three legged diagram. It is convenient to color
	the legs of the diagram to keep track of the extra factors between the $3j$ symbols and the Clesbsh-Gordan symbols. $iii)$ The diagram with a black dot 
	corresponds to the components of the singlet.}\label{fig:Diagrams}
\end{figure}

It is convenient to introduce a diagrammatic presentation of the different tensors appearing above, as otherwise the 
notation becomes quickly very cumbersome. This presentation also makes clear which manipulations are being done with the 
different expressions. The building blocks for the diagrammatic presentation are presented in Fig. \ref{fig:Diagrams}.
The concatenation of two objects by joining their lines corresponds to sum over all the possible indices 
of the lines at the concatenation point. The concatenation of a bra and a ket corresponds to taking their inner product.

Using this presentation, the groundstate $|\mathcal{G}\rangle$ is simply
\begin{equation}
		 \includegraphics[width=0.8\linewidth]{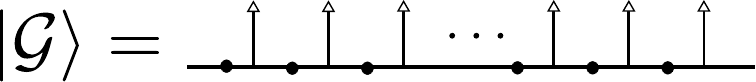},
\end{equation}
where we omit the coloring of the Clebsch-Gordan coefficients as is clear how to contract them.
The state $|\mathcal{G}\rangle$ can be thought as a very simple one dimensional lattice gauge theory, where each horizontal line
carries a $SU(2)$ singlet (trivial representation). Each vertical line carries a spin $S$ representation. Attached to this 
representation is a state that acts on the physical Hilbert space.
The one dimensional transfer matrix $G$, defined in (\ref{TM_2}), has the following presentation
\begin{equation}\label{pres_trans_matrix}
		 \includegraphics[width=0.8\linewidth]{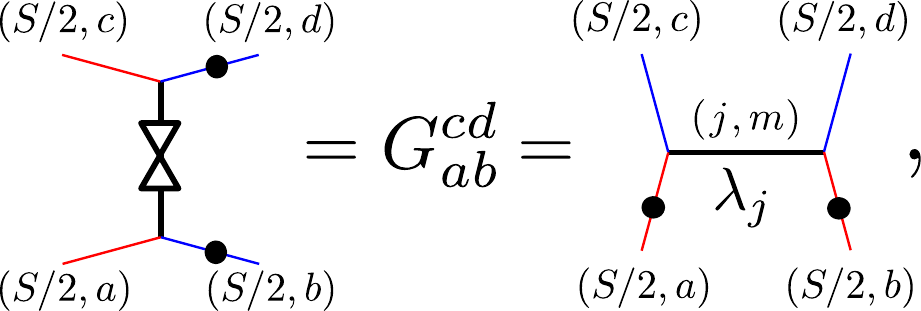}
\end{equation}
where the values $(j,m)$ of the internal line on the right diagram are being summed over, with weight $\lambda_j$.

\section{Reduced density matrix}

We are interested in the partial density matrix obtained by tracing subsystems of the original pure ground state.
We partition the system into five distinct consecutive regions, of lengths $L_1,L_A,L_2,L_B$, and $L_3$. 
Tracing out the spin degrees of freedom in the first, third and last regions, we obtain a mixed state described
by the partial density matrix $=\rho_{AB}={\rm tr}_{1,2,3}(|\mathcal{G}\rangle\langle\mathcal{G}|)$. This density matrix, for any boundary condition, corresponds to the 
diagram
\begin{equation}
		 \includegraphics[width=\linewidth]{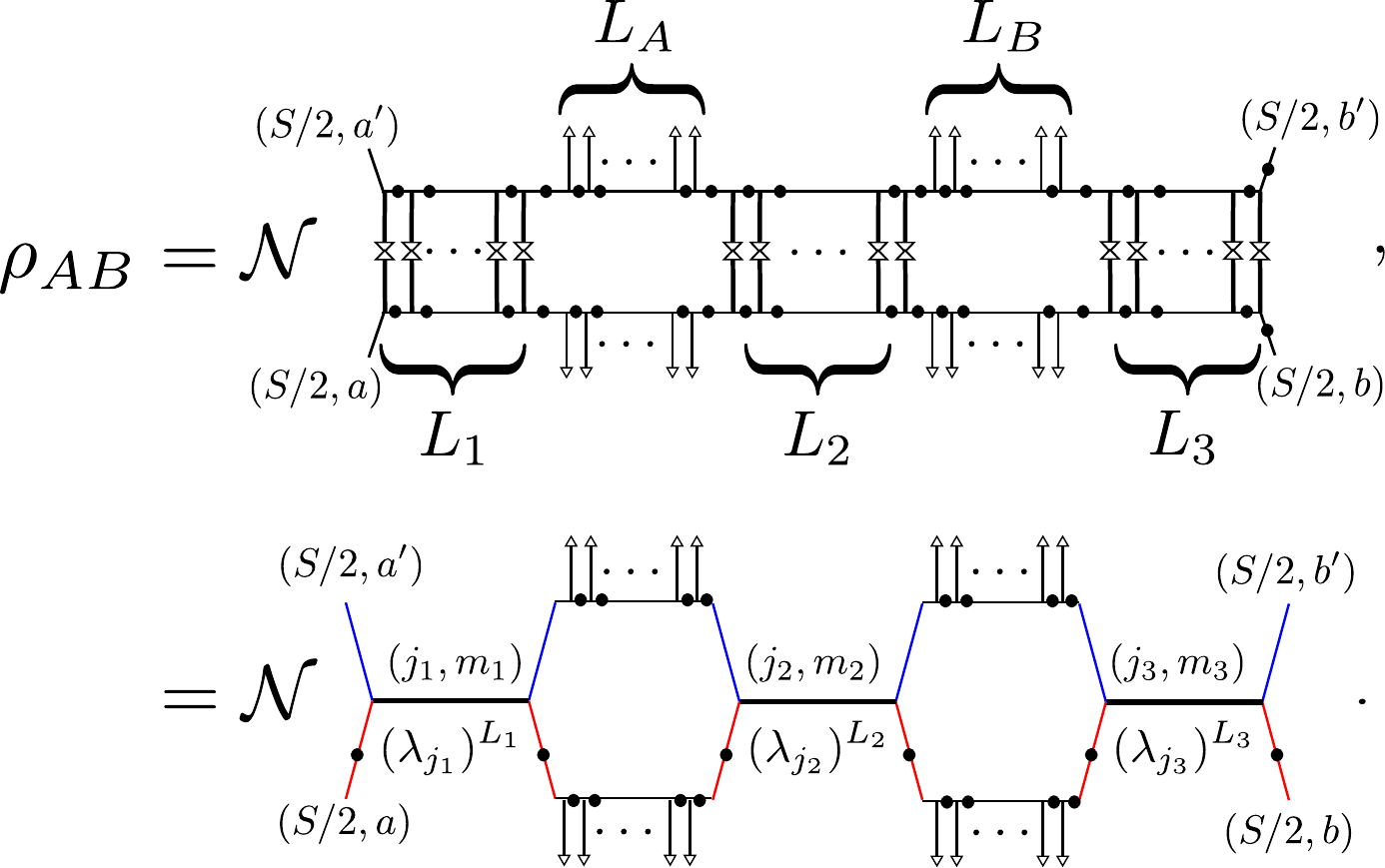}
	%\caption{Partial density matrix $\rho_{AB}$.}\label{fig:PartialDM}
\end{equation}

\noindent Here $\mathcal{N}$ is a normalization constant such that ${\rm tr}\rho_{AB}=1$. Internal lines with no
explicit label carry an $S/2$ representation.
Different boundary conditions are 
implemented by different contractions of the outermost tensors. The lower diagram is obtained by repeated use of eq. (\ref{TM_2})
on the upper diagram. The colors in the upper diagram, corresponding to the original MPS presentation, are omitted.
In the bottom diagram, and in the rest of the text, the different colors of the legs in the Clebsch-Gordan coefficients are explicitly shown.

The partial density matrix $\rho_{AB}$ is an operator that acts in the Hilbert space of the $A$ and $B$ subsystems. 
It is convenient to introduce a orthogonal basis on the $A,B$ subsystems. We choose the orthogonal basis (See appendix \ref{AppC})
\begin{equation}
		\vcenter{\hbox{ \includegraphics[width=0.8\linewidth]{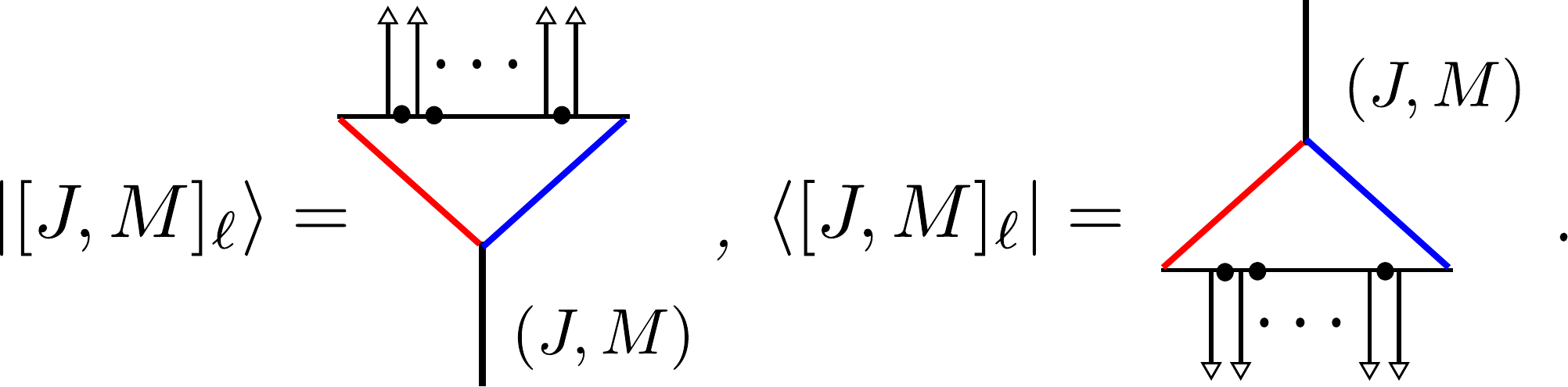}}}
	%\caption{Partial density matrix $\rho_{AB}$.}\label{fig:PartialDM}
\end{equation}

In this basis, the density matrix operator is given by the diagram
\begin{equation}
		\vcenter{\hbox{ \includegraphics[width=0.9\linewidth]{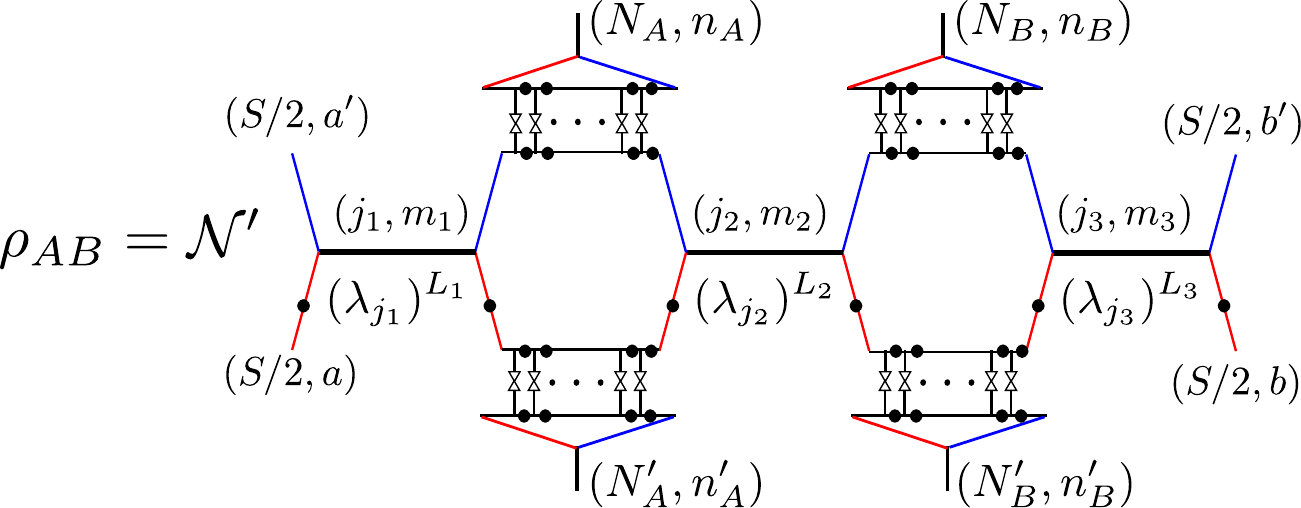}}},
	%\caption{Partial density matrix $\rho_{AB}$.}\label{fig:PartialDM}
\end{equation}

\noindent where the normalization $\mathcal{N}'$ is explicitly
\begin{equation}
 \mathcal{N}'=\frac{\mathcal{N}}{\sqrt{\eta_{N_A}^{(L_A)}\eta_{N'_A}^{(L_A)}\eta_{N_B}^{(L_B)}\eta_{N'_B}^{(L_B)}}},
\end{equation}
with $\eta_{J_\alpha}^{(L_\alpha)}=\sum_{k=0}^S(\lambda_k)^{L_\alpha}F_{J_\alpha,k}$,
and $\alpha=A,B$. Using the crossing symmetry relation (\ref{TM_2},\ref{pres_trans_matrix}) (see also appendix \ref{appendixB}), the density matrix
$\rho_{AB}$ becomes
\begin{figure}[h!]
		 \includegraphics[width=0.9\linewidth]{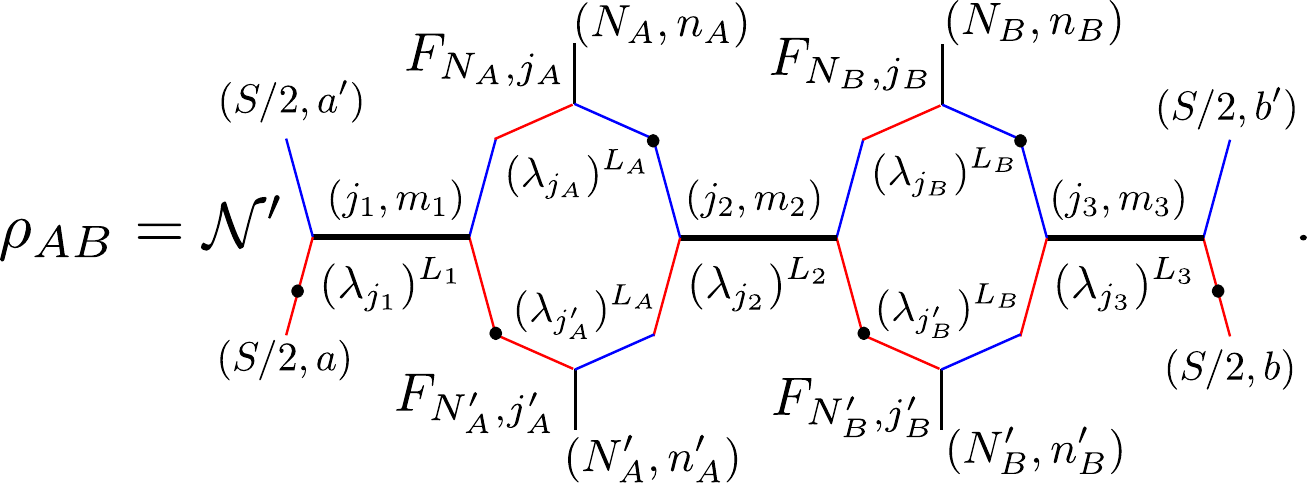}
	%\caption{Partial density matrix $\rho_{AB}$.}\label{fig:PartialDM}
\end{figure}

By applying a series of recolorings and $F$ moves (Appendix \ref{appendixA}) the partial density matrix can be written in 
the standard basis
\begin{equation}\label{PDM_GBC}
	 \rho_{AB}=\sum_{\substack{P,Q\\j_1j_2j_3}}X_{N_A'N_B';j_1j_2j_3}^{N_AN_B;PQ}\vcenter{\hbox{\includegraphics[width=0.5\linewidth]{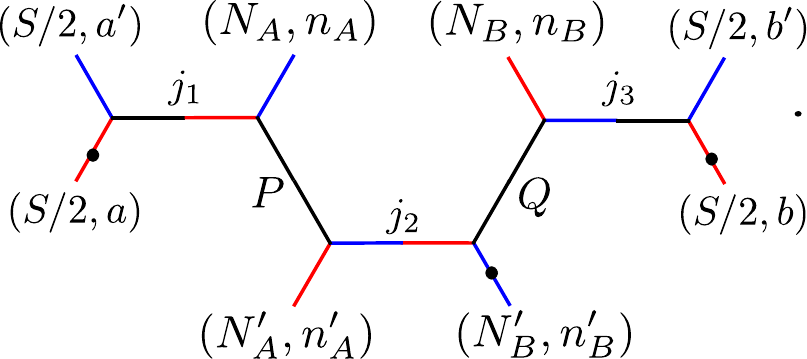}}}
\end{equation}
Here $X_{n_3n_4;j_1j_2j_3}^{n_1n_2;PQ}$ is a combination of
$F$-symbols, given explicitly in appendix \ref{AppendixD}.
Note that each line in the diagram (\ref{PDM_GBC}) carries a $SU(2)$ representation label, and
the vertex where three lines meet corresponds to a Clebsch-Gordan symbol. An object with these characteristics
(i.e a lattice with representations associated to each link, and associators when three or more representation meet) is known in 
the mathematical literature as a state sum \cite{Kirillov2011}. In physics, such an object corresponds to a very simple Levin-Wen
model \cite{Levin2005} (without long-range order).

\subsection*{Periodic boundary conditions and the thermodynamic limit}

To make further progress, we need to choose some boundary conditions.
Assuming periodic boundary conditions (PBC) amounts for contracting the outermost
tensors in the diagram above. 
In the standard basis, after a series of $F$-moves and recolorings, the reduced density matrix becomes finally
\begin{equation}\label{DM_PBC}
\rho_{AB}=\sum_{R=|N_A-N_B|}^{N_A+N_B}\Gamma_{N'_AN'_B}^{N_AN_B}(R)\vcenter{\hbox{\includegraphics[height=2.5cm]{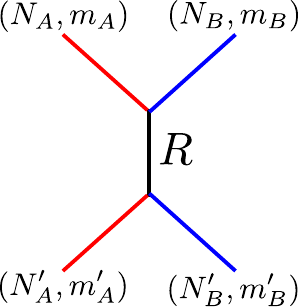}}}.
\end{equation}
An expression of $\Gamma$ is given in appendix \ref{AppendixD} in terms
of $6j$-symbols. From this expression, the eigenvalues $\Lambda^{(\alpha,R)}$ of the density matrix can be obtained by solving 
the eigenvalue equation
\begin{equation}
 \sum_{n_3n_4}\Gamma_{n_3n_4}^{n_1n_2}(R)e_{n_3n_4}^{(\alpha,R)}=\Lambda^{(\alpha,R)}e_{n_1n_2}^{(\alpha,R)}.
\end{equation}

In the thermodynamic limit of $L_1+L_3\rightarrow\infty$, the eigenvalues and eigenvectors 
of $\rho_{AB}$ can be obtained explicitly as $\rho_{AB}$ becomes
\begin{equation}
 \rho_{AB}=\sum_R\Lambda^R_{N_AN_B}\delta_{N_A'}^{N_A}\delta_{N_B'}^{N_B}\vcenter{\hbox{\includegraphics[height=2.5cm]{D1.pdf}}},
\end{equation}
where $\delta_{a}^b$ is a Kronecker delta function.
The eigenvalues of $\rho_{AB}$, labeled by $n,m\in[0,S]$ and 
$R\in[|n-m|,n+m|$ are
\begin{eqnarray}\nonumber
\Lambda_{NM}^R&=&\sum_{j=0}^S\left({\lambda_j}\right)^{L_2}\SixJ{M}{N}{N}{M}{j}{R}\SixJ{N}{\frac{S}{2}}{N}{\frac{S}{2}}{j}{\frac{S}{2}}
\SixJ{\frac{S}{2}}{M}{\frac{S}{2}}{M}{j}{\frac{S}{2}}\\
&\times&(-1)^{R+j}\frac{(2N+1)(2M+1)\eta^{(L_A)}_N\eta^{(L_B)}_M}{S+1}.
\end{eqnarray}

\subsection*{Spin $S/2$ at the boundaries}

The expression for the reduced density matrix simplifies notably in the case when at the two ends of the chain a spin $S/2$ 
particle is located. This corresponds to contracting the upper and lower outermost tensor in (\ref{PDM_GBC}). After some 
straightforward manipulations we have
\begin{equation}\label{spinhalfboundary}
\rho_{AB}=\sum_{j_2,R}Y_{N_A'N_B';j_2R}^{N_AN_B}\vcenter{\hbox{\includegraphics[width=0.25\linewidth]{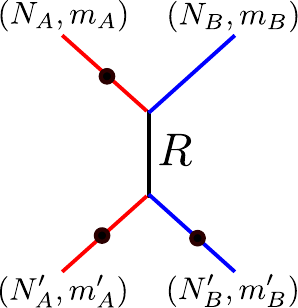}}}.
\end{equation}
The tensor $Y_{N_A'N_B';j_2R}^{N_AN_B}$ is given explicitly in (\ref{S/2BCtensor}).

\section{Negativity for the mixed system of 2 blocks}\label{N2systems}

The negativity of a the subsystem $A$ is defined as the sum of the negative eigenvalues of the PTDM respect to the subsystem $A$
\begin{equation}
 {\rm Neg}(\rho_{AB})=\frac{||\rho_{AB}^{T_A}||-1}{2}=\sum_i\frac{|r_i|-r_i}{2},
\end{equation}
with $r_i$ the eigenvalues of $\rho_{AB}^{T_A}$. The norm of the operator $||O||$ is defined as $ ||O||={\rm Tr}\sqrt{O^\dagger O}.$
Following the same steps that lead us to the density matrix (\ref{PDM_GBC}), we find
\begin{equation}\label{DM_ABC_Trans}
\rho_{AB}^{T_A}=\sum_{\substack{P,Q\\j_1j_2j_3}}X_{N_AN_B';j_1j_2j_3}^{N_A'N_B;PQ}\vcenter{\hbox{\includegraphics[height=2cm]{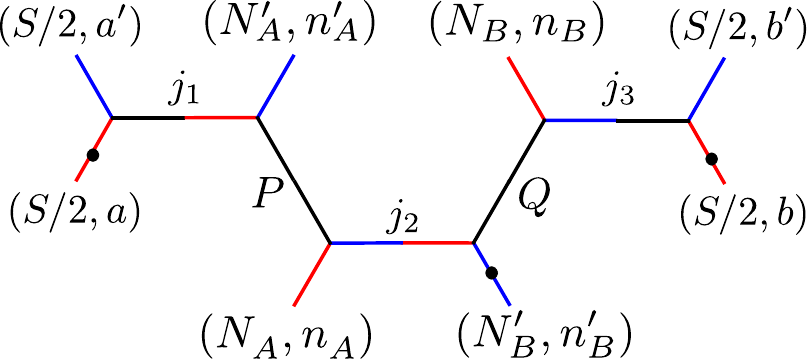}}}.
\end{equation}
where the states $(N_A,m_a)$ and $(N_A',m_a')$ have been transposed with respect to (\ref{PDM_GBC}).

\begin{figure}
 \includegraphics[width=\linewidth]{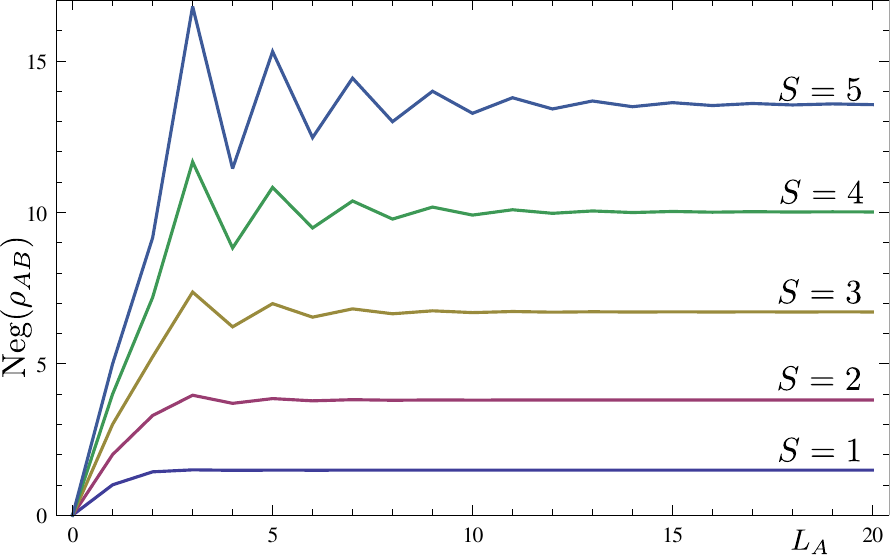}
 \caption{Negativity of the PTDM as function of block length $L_A$, for fixed size of block $B$, $(L_B=2)$. Different
 curves represent different values of spin $S$. The continuous line is just a guide to the eye.}\label{fig:Neg}
\end{figure}

The partial transposed density matrix $\rho_{AB}^{T_A}$ (with PBC) is
\begin{equation}\label{DM_PBCPT}
\rho_{AB}^{T_A}=\sum_{R=|N_A-N_B|}^{N_A+N_B}{\Gamma}_{N_AN'_B}^{N_A'N_B}(R)\vcenter{\hbox{\includegraphics[height=2.5cm]{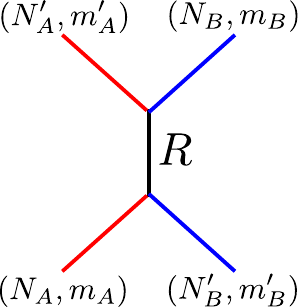}}},
\end{equation}
In the case of two adjacent blocks $A$ and $B$, where $L_1=L_2=L_3$ we find after a straightforward calculation that the operator $\rho_{AB}^{T_A}$ has the simple form
\begin{equation}\label{PTDM_1}
 U\rho_{AB}^{T_A}U^\dagger=\sum_{\substack{N,M\\n,m}}\frac{\sqrt{\eta_N^{(L_A)}\eta_N^{(L_B)}\eta_M^{(L_A)}\eta_M^{(L_B)}}}{\mathcal{N}_{\rm PBC}}|^N_n,\,^M_m\rangle\langle^M_m,\,^N_n|,
\end{equation}
where $|^N_n\rangle\equiv|N,n\rangle$ is a state of total spin $N$ and $S_z$ projection $n$. The unitary transformation $U$ is
defined by $U|^N_m\rangle=(-1)^n|^N_{-n}\rangle$. The normalization for a system of total length $L_T$ and PBC is
\begin{equation}
 \mathcal{N}_{\rm PBC}=\sum_{j=0}^S(2j+1)(\lambda_j)^{L_T}.
\end{equation}
The eigenvectors and eigenvalues $(e,\lambda)$ of the operator  (\ref{PTDM_1}) are
\begin{eqnarray}\label{eigenvalues1}
 &\left(|^N_n,^N_n\rangle, \frac{\eta_N^{(L_A)}\eta_N^{(L_B)}}{\mathcal{N}_{PBC}}\right),\\\label{eigenvalues2}
 &\left(\frac{|^N_n,^M_m\rangle\pm|^M_m,^N_n\rangle}{\sqrt{2}}, \frac{\pm\sqrt{\eta_N^{(L_A)}\eta_N^{(L_B)}\eta_M^{(L_A)}\eta_M^{(L_B)}}}{\mathcal{N}_{PBC}}\right),
\end{eqnarray}
where (\ref{eigenvalues2}) is valid for $N\neq M$ and $n\neq m$.
Taking into account the degeneracies of the eigenvalues, we find the negativity to be
\begin{widetext}
\begin{equation}\label{Neg_adjacent}
 {\rm Neg}(\rho_{AB})=\mathcal{N}_{\rm PBC}^{-1}\sum_{N=0}^S(2N+1)\sqrt{\eta_N^{(L_A)}\eta_N^{(L_B)}}\left[N\sqrt{\eta_N^{(L_A)}\eta_N^{(L_B)}}+\sum_{M=N+1}^S(2M+1)\sqrt{\eta_M^{(L_A)}\eta_M^{(L_B)}}\right],
\end{equation}
\end{widetext}
valid for a periodic chain of total length $L_T=L_A+L_B$ with $L_{A(B)}$ is the size of the block $A\,(B)$. For future reference we write
explicitly 
\begin{equation}
 \eta_{N}^{(L)}=\sum_{j=0}^S(2j+1)(-1)^{N+j+S}(\lambda_{j})^{L}\SixJ{S/2}{S/2}{S/2}{S/2}{j}{N}.
\end{equation}
with $\lambda_j$ given by eq. (\ref{lambdaj}).
Note that (\ref{Neg_adjacent}) is symmetric under an interchange of $A$ and $B$, a trivial consequence of $\rho_{AB}^{T_A}=\rho_{AB}^{T_B}$.
Some particular simple cases are 
\begin{itemize}
 \item $L_B=1$ for any $L_A\longrightarrow{\rm Neg}(\rho_{AB})=S$.
 \item $L_A,L_B\gg =1\longrightarrow{\rm Neg}(\rho_{AB})=\frac{S(S+2)}{2}.$
\end{itemize}
Other generic cases are plotted in Fig. \ref{fig:Neg}.

\subsection*{General Block separation and LOCC}

The negativity found in (\ref{Neg_adjacent}) is valid just for adjacent blocks $(L_1=L_2=L_3=0)$.
From the exact density matrix (\ref{DM_PBC}) for PBC and (\ref{spinhalfboundary}) for spin $S/2$ particles in the boundary,
we can compute the negativity for any fixed value of $S, L_A,L_B,L_1,L_2$ and $L_3$ by exact (numerical) diagonalization. 
In the case of $S=1,2,3$ and PBC we that whenever $L_1+L_3=L_2=1$, the negativity vanishes. 
For general spin $S$ and separations $L_1+L_3,L_2\ge1$, we conjecture the following
\begin{itemize}
 \item {\bf Conjecture}: The negativity of a bipartite system of two blocks $A$ and $B$ in the one dimensional spin $S$ VBS 
 state vanishes for blocks with no common boundary.
\end{itemize}
In Ref. \cite{Santos2012} we proved this conjecture for spin $S=1$. Although a proof for a given value of $S$ can be obtained by a direct 
computation of the negativity for different lengths, for arbitrary value of $S$ this is impracticable.
We can further restrict the direct computation to just
$L_1+L_3=L_2=1$ (for spin $S/2$ boundary conditions). As the VBS state can be constructed inductively by series of local 
operations (projection of two spin $S/2$ particles in each physical site onto the symmetric subspace, plus 
an antisymmetrization of spin $S/2$ particles on consecutive physical sites), a vanishing negativity ${\rm Neg}(\rho_{AB})$
for a particular length $\ell$ implies that ${\rm Neg}(\rho_{AB})=0$ for all $L\geq \ell$, due to the monotonicity of the 
negativity under LOCC. We have indeed verified that this is the case for $S=2,3$.

\section{Conclusion}
The AKLT model represents a simple interacting many-body system of interest for both the condensed matter and 
quantum information community. It realizes the Haldane phase, a gapped phase with symmetry protected topological order.
Its VBS groundstate, also serves as a resource for measurement based quantum information. Although being an interacting theory, 
the VBS groundstate can be written exactly by using simple MPS. The matrices in these MPS are naturally Clebsch-Gordan
coefficients between representations of spin $S$ and spin $S/2$ for a chain of spin $S$ particles. Using the representation
theory of the $SU(2)$ algebra, we study the bipartite entanglement of the mixed state density matrix obtained by tracing out 
different sections of the pure state density matrix. 
In particular, we managed to calculate negativity between two blocks of spins analytically, in the case when the blocks are
adjacent. Our results are exact for any value of the spin $S$ as for any length of the blocks, in a chain with periodic boundary
conditions. For non adjacent blocks, we checked that the negativity vanishes for $S=1,2,3$, and conjectured that the negativity 
vanishes for any value of the spin $S$. We think that vanishing negativity for two non-touching blocks is unique characterization 
of VBS states. For generic short-range entangled states, we expect the negativity to decay exponentially with the separation 
between the blocks, with a characteristic decay length proportional to the correlation length.
The methods used here allow the computation of multipartite entanglement density matrix, where the $SU(2)$ symmetry is manifest
at each stage. We expect that the methods presented here, which rely on the associativity of the tensor product of 
representations, can be extended to the study of ground state properties in string-net modes, where the categorical data 
in such models satisfies similar relations to the ones used in this work.

\textbf{Acknowledgments}
R.S. acknowledges F. Duarte and the Physical and Theoretical Chemistry Laboratory at Oxford University, where substantial 
part of this work was performed. V.K. acknowledges productive atmosphere of Simons Center for Geometry and Physics.

\smallskip

\appendix

\onecolumngrid
\section{Identities of Clebsch-Gordan coefficients}\label{appendixA}

The Clebsch-Gordan coefficients satisfy a series of identities. Casted in the diagrammatic presentation, they read

\begin{eqnarray}
		&\vcenter{\hbox{\includegraphics[width=0.7\linewidth]{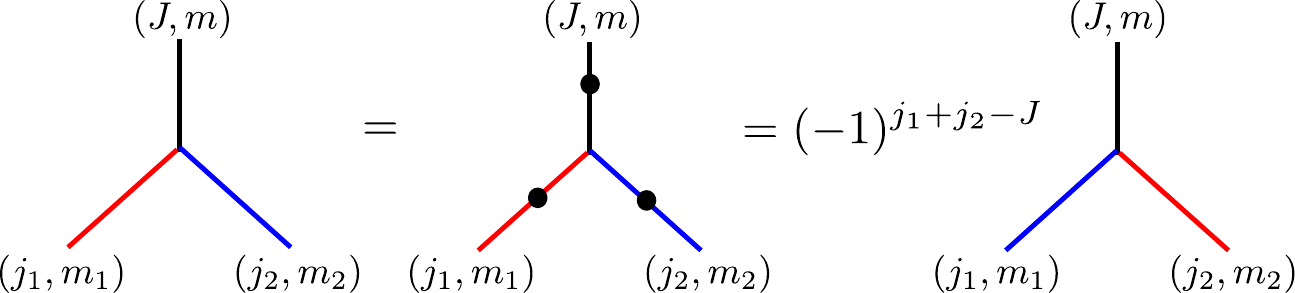}}},\\
		&\vcenter{\hbox{\includegraphics[width=0.9\linewidth]{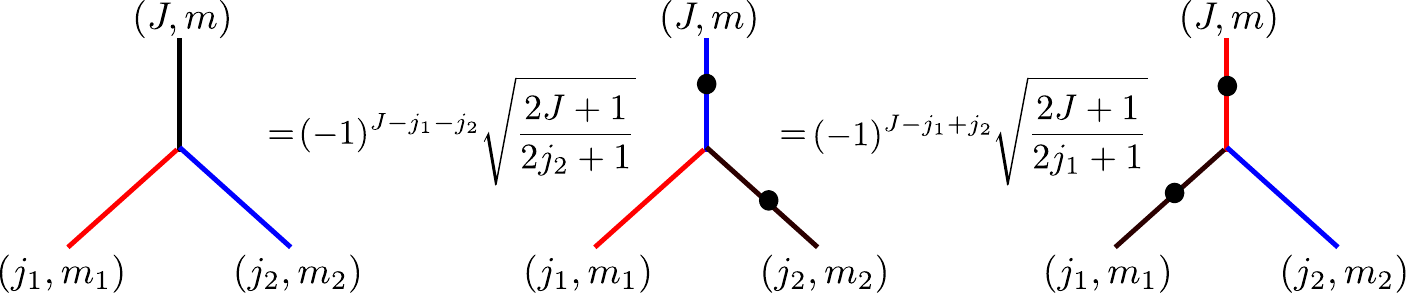}}}.
\end{eqnarray}
The orthogonality of the Clebsch-Gordan coefficients reads diagrammatically
\begin{equation}
 \vcenter{\hbox{\includegraphics[width=0.7\linewidth]{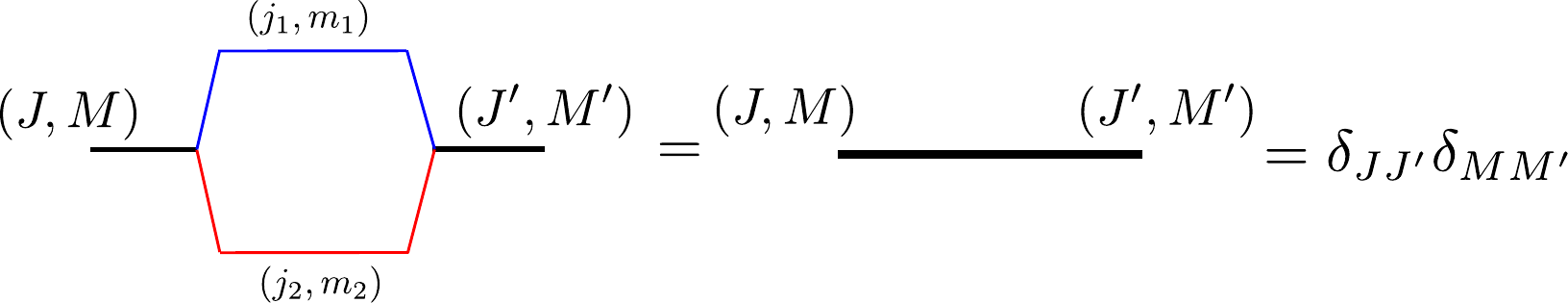}}}.
\end{equation}

\section{$F$-moves }\label{appendixB}
\begin{figure}[h!]
		 \includegraphics[width=0.5\linewidth]{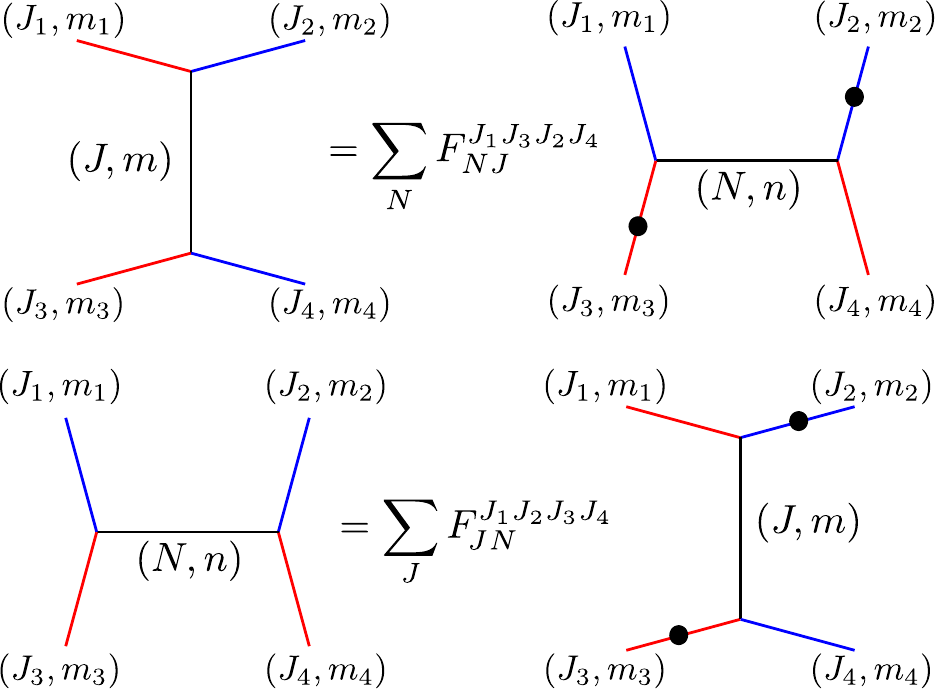}
	\caption{Recoupling of Clebsch-Gordan coefficients.}\label{fig:F_matrix}
\end{figure}

%%%%%%%%%%%%%%%%%
In order to compute the eigenvalues of the transfer matrix $\tilde{T}$, we make use of the SU($2$) structure of $\tilde{T}$. This 
matrix corresponds to the contraction of Clebsch-Gordan symbols, so we can use the $F-$matrix (Racah coefficients 
\cite{messiah2014quantum}) to recouple the coefficients. The recoupling is expressed in Fig. \ref{fig:F_matrix}

The $F-$matrix is related to the $6j$-symbol by
\begin{equation}
 F^{J_1J_2J_3J_4}_{NJ}=(-1)^{J_1-2J_3-J_4+N-J}(2J+1)\SixJ{J_1}{J_4}{J_2}{J_3}{N}{J}.
\end{equation}

\section{Orthogonal basis for arbitrary length}\label{AppC}
The orthogonality of the $|[J,M]_\ell\rangle$ basis can be already verified by manipulating the corresponding diagram
\begin{figure}[h!]
		 \includegraphics[width=0.6\linewidth]{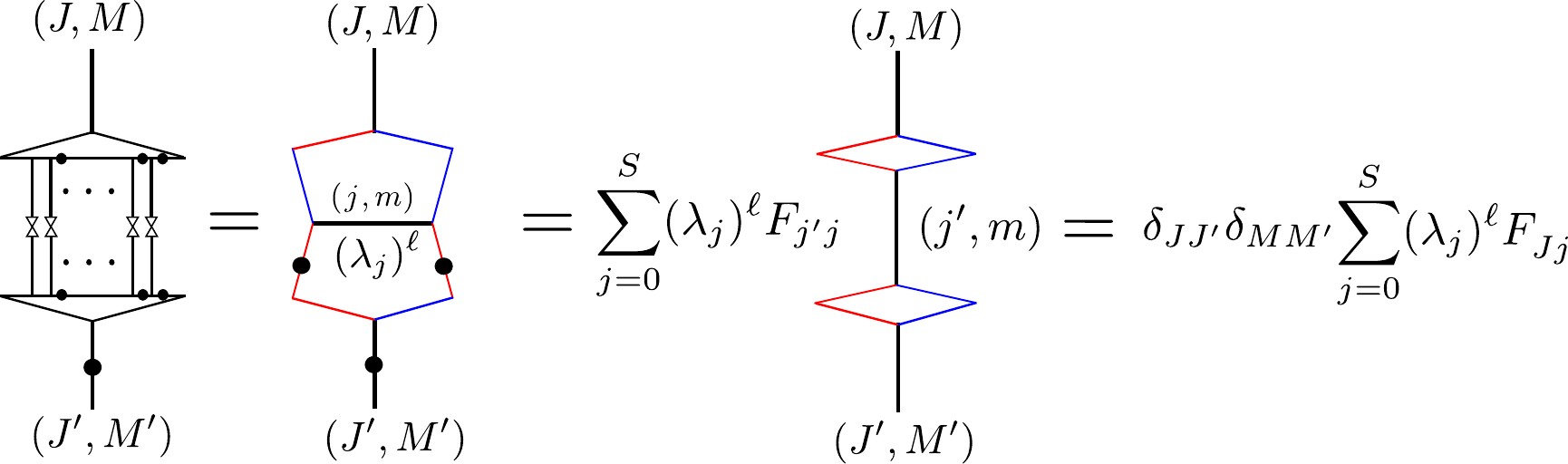}
	%\caption{Partial density matrix $\rho_{AB}$.}\label{	fig:PartialDM}
\end{figure}

\noindent where $F_{j'j}\equiv F_{j'j}^{S/2, S/2, S/2, S/2}$ is the $F-$symbol introduced previously.

\section{Explicit Tensors of reduced density matrix}\label{AppendixD}

The $X$ tensor that determines the partial density matrix with general boundary conditions (\ref{PDM_GBC}) is 

\begin{eqnarray}\label{X_tensorGBC}\nonumber
X_{n_3n_4;j_1j_2j_3}^{n_1n_2;PQ}=\mathcal{N}\sqrt{2j_2+1}(-1)^{n_1+n_2+j_2+j_3}\prod_{k=1}^4\frac{\sqrt{(2n_k+1)\eta_{n_k}}}{S+1}\prod_{p=1}^3{\sqrt{2j_p+1}(\lambda_{j_p})^{L_p}}
 F_{P\frac{S}{2}}^{n_1j_1\frac{S}{2}\frac{S}{2}}F_{P\frac{S}{2}}^{\frac{S}{2}\frac{S}{2}n_3j_2}F_{Q\frac{S}{2}}^{n_2j_3\frac{S}{2}\frac{S}{2}}F_{Q\frac{S}{2}}^{\frac{S}{2}\frac{S}{2}n_4j_2}.
\end{eqnarray}

The $Y$ tensor that appears in $\rho_{AB}$ with spin $S/2$ boundary conditions (\ref{spinhalfboundary}) is 

\begin{eqnarray}\label{S/2BCtensor}\nonumber
Y_{n_3n_4;j_2R}^{n_1n_2}&=&(\lambda_{j_2})^{L_2}(-1)^{n_1+n_3}\prod_{k=1}^4\sqrt{\frac{(2n_k+1)\eta_{n_k}}{S+1}}
 F_{n_1\frac{S}{2}}^{\frac{S}{2}\frac{S}{2}n_3j_2}F_{n_2\frac{S}{2}}^{\frac{S}{2}\frac{S}{2}n_4j_2}F_{Rj_2}^{n_4n_3n_2n_1},\\
 &=&(\lambda_{j_2})^{L_2}(-1)^{n_1+n_3+R+j_2}(2j_2+1)\SixJ{n_1}{\frac{S}{2}}{n_3}{\frac{S}{2}}{j_2}{\frac{S}{2}}
 \SixJ{\frac{S}{2}}{n_2}{\frac{S}{2}}{n_4}{j_2}{\frac{S}{2}}
 \SixJ{n_2}{n_3}{n_4}{n_1}{j_2}{R}\prod_{k=1}^4\sqrt{(2n_k+1)\eta_{n_k}}.
\end{eqnarray}

The $\Gamma$ tensor appearing in the partial density matrix with periodic boundary conditions (\ref{DM_PBC})
is given by

\begin{eqnarray}\label{GammaTensorPBC}\nonumber
\Gamma_{n_3n_4}^{n_1n_2}(r)=\mathcal{N_{\rm PBC}}(-1)^{n_1+n_2}\prod_{k=1}^4\sqrt{(2n_k+1)\eta_{n_k}}\sum_{p,q,j_1,j_2}(2p+1)(2q+1)(2j_1+1)(2j_2+1)(-1)^{p+q+j_1+j_2}\\
\times(\lambda_{j_1})^{L_1+L_3}(\lambda_{j_2})^{L_2}\SixJ{q}{n_3}{p}{n_4}{r}{j_2}\SixJ{n_1}{\frac{S}{2}}{j_1}{\frac{S}{2}}{p}{\frac{S}{2}}
\SixJ{n_2}{\frac{S}{2}}{j_1}{\frac{S}{2}}{q}{\frac{S}{2}}\SixJ{q}{n_1}{p}{n_2}{r}{j_1}
\SixJ{n_3}{\frac{S}{2}}{j_2}{\frac{S}{2}}{p}{\frac{S}{2}}\SixJ{n_4}{\frac{S}{2}}{j_2}{\frac{S}{2}}{q}{\frac{S}{2}}.
\end{eqnarray}

The $\tilde{\Gamma}$ tensor that appears in $\rho_{AB}^{T_A}$ is in turn

\begin{eqnarray}\label{GammaTensorPBCPT}\nonumber
\Gamma_{n_3n_4}^{n_1n_2}(r)=\mathcal{N_{\rm PBC}}(-1)^{n_1+n_2}\prod_{k=1}^4\sqrt{(2n_k+1)\eta_{n_k}}\sum_{p,q,j_1,j_2}(2p+1)(2q+1)(2j_1+1)(2j_2+1)(-1)^{p+q}(\lambda_{j_1})^{L_1+L_3}\\
\times(\lambda_{j_2})^{L_2}\SixJ{q}{n_3}{p}{n_4}{r}{j_2}\SixJ{n_1}{\frac{S}{2}}{j_1}{\frac{S}{2}}{p}{\frac{S}{2}}
\SixJ{n_2}{\frac{S}{2}}{j_1}{\frac{S}{2}}{q}{\frac{S}{2}}\SixJ{q}{n_1}{p}{n_2}{r}{j_1}
\SixJ{n_3}{\frac{S}{2}}{j_2}{\frac{S}{2}}{p}{\frac{S}{2}}\SixJ{n_4}{\frac{S}{2}}{j_2}{\frac{S}{2}}{q}{\frac{S}{2}}.
\end{eqnarray}

\twocolumngrid

\end{document}